# Fermi surface nesting in several transition metal dichalcogenides.

D. S. Inosov,[1] V. B. Zabolotnyy,[1] D. V. Evtushinsky,[1] A. A. Kordyuk,[1,2]
B. Büchner,[1] R. Follath,[3] H. Berger,[4] and S. V. Borisenko[1]

[1]*Institute for Solid State Research, IFW Dresden, P.O. Box 270116, D-01171 Dresden, Germany.*
[2]*Institute of Metal Physics of National Academy of Sciences of Ukraine, 03142 Kyiv, Ukraine.*
[3]*BESSY GmbH, Albert-Einstein-Strasse 15, 12489 Berlin, Germany.*
[4]*Institut de Physique de la Matière Complexe, EPFL, 1015 Lausanne, Switzerland.*

By means of high-resolution angle resolved photoelectron spectroscopy (ARPES) we have studied the fermiology of 2H transition metal dichalcogenide polytypes TaSe$_2$, NbSe$_2$, and Cu$_{0.2}$NbS$_2$. The tight-binding model of the electronic structure, extracted from ARPES spectra for all three compounds, was used to calculate the Lindhard function (bare spin susceptibility), which reflects the propensity to charge density wave (CDW) instabilities observed in TaSe$_2$ and NbSe$_2$. We show that though the Fermi surfaces of all three compounds possess an incommensurate nesting vector in the close vicinity of the CDW wave vector, the nesting and ordering wave vectors do not exactly coincide, and there is no direct relationship between the magnitude of the susceptibility at the nesting vector and the CDW transition temperature. The nesting vector persists across the incommensurate CDW transition in TaSe$_2$ as a function of temperature despite the observable variations of the Fermi surface geometry in this temperature range. In Cu$_{0.2}$NbS$_2$ the nesting vector is present despite different doping level, which lets us expect a possible enhancement of the CDW instability with Cu-intercalation in the Cu$_x$NbS$_2$ family of materials.

PACS numbers: 71.45.Lr 79.60.-i 71.18.+y 74.25.Jb

**Introduction**

In 1955 Peierls [1] suggested that in one-dimensional metals a spontaneous formation of periodic lattice distortions (PLD) and charge density waves (CDW) can be energetically favorable under certain conditions. Since then, CDW formation has been experimentally observed in many anisotropic compounds, such as transition metal chalcogenides [2–6]. This kind of symmetry breaking, which happens upon cooling at a certain transition temperature $T_{\rm CDW}$, is known as a Peierls phase transition. The physical mechanisms of CDW formation are now well understood and are generally known to be determined in particular by the Fermi surface geometry [1, 7–9], though some aspects of CDW formation in two-dimensional metals, including its possible relation to the problem of high-$T_{\rm c}$ superconductivity, are still actively discussed [4–6, 8, 10–21].

In its simplest form, the instability condition for the formation of CDW/PLD in an electronic system is given by [7]

$$4\bar{\eta}_{\bf q}^2/\hbar\omega_{\bf q} - 2\bar{U}_{\bf q} + \bar{V}_{\bf q} \geq 1/\chi_{\bf q}, \tag{1}$$

where $\bar{U}_{\bf q} = \langle {\bf k}+{\bf q}\,{\bf k'}|\hat{U}|{\bf k'}+{\bf q}\,{\bf k}\rangle$ and $\bar{V}_{\bf q} = \langle {\bf k}+{\bf q}\,{\bf k'}|\hat{V}|{\bf k}\,{\bf k'}+{\bf q}\rangle$ are the direct and exchange Coulomb interactions in the local approximation [22], $\bar{\eta}_{\bf q}$ is the local electron-phonon interaction, and $\chi_{\bf q} = \sum_{\bf k}[n_{\rm F}(\epsilon_{\bf k}) - n_{\rm F}(\epsilon_{\bf k+q})]/(\epsilon_{\bf k} - \epsilon_{\bf k+q})$ is the real part of the bare spin susceptibility (Lindhard function) at $\omega \to 0$, which can be successfully evaluated from the ARPES data [23]. Here by $n_{\rm F}(\epsilon) = 1/[\exp(\epsilon/k_{\rm B}T) + 1]$ we denote the Fermi-Dirac distribution function. Note that the imaginary part of $\chi_{\bf q}$ vanishes in the static limit.

From Eq. (1) one sees that if the electron-phonon interaction is strong enough for the left part of the inequality to be positive, a divergence or a strong peak in $\chi_{\bf q}$ at a particular wave vector ${\bf q}$ would lead to the CDW/PLD instability. The phase transition would be then preceded by the softening of a phonon mode, until it "freezes" at $T_{\rm CDW}$, giving rise to the PLD with the same (or similar) wave vector ${\bf q}$. Appearance of such a divergence in the static susceptibility we will call *nesting*. In the simplest scenario, such sharp peak will arise if the Fermi surface possesses parallel fragments such that many pairs of electronic states can be connected by the same wave vector ${\bf q}$, which results in an enhancement of the susceptibility at this vector.

It is a long-standing argument, however, whether such simple mechanism of Fermi surface instabilities underlies the CDW formation in transition metal dichalcogenides, such as TaSe$_2$ and NbSe$_2$, which are the subject of this letter. In some of the earlier studies the existence of necessary nesting conditions in transition metal chalcogenides was questioned [18, 24–26], and some alternative mechanisms of CDW instability were proposed [8, 10, 14, 16]. We find several instability scenarios proposed in the literature: (i) simple Fermi surface nesting [27], which in some studies was considered too weak to be responsible for the instability [28], (ii) nesting of the van Hove singularities (saddle points) [8, 29, 30], and (iii) combination of the two: partial nesting of the FS with the saddle band [31].

To clarify the role of simple nesting as the driving force of the CDW instabilities, we performed high resolution measurements of several transition metal dichalcogenides using modern angle-resolved photoelectron spectroscopy, which let us accurately determine the Fermi surface geometries and assess their nesting properties and their variations with temperature. As will be shown in the following, our results not only support the Fermi surface nesting scenario of CDW formation, but also reveal new aspects of nesting geometry: incommensurability of the nesting vector and its universality among several transition metal dichalcogenides.

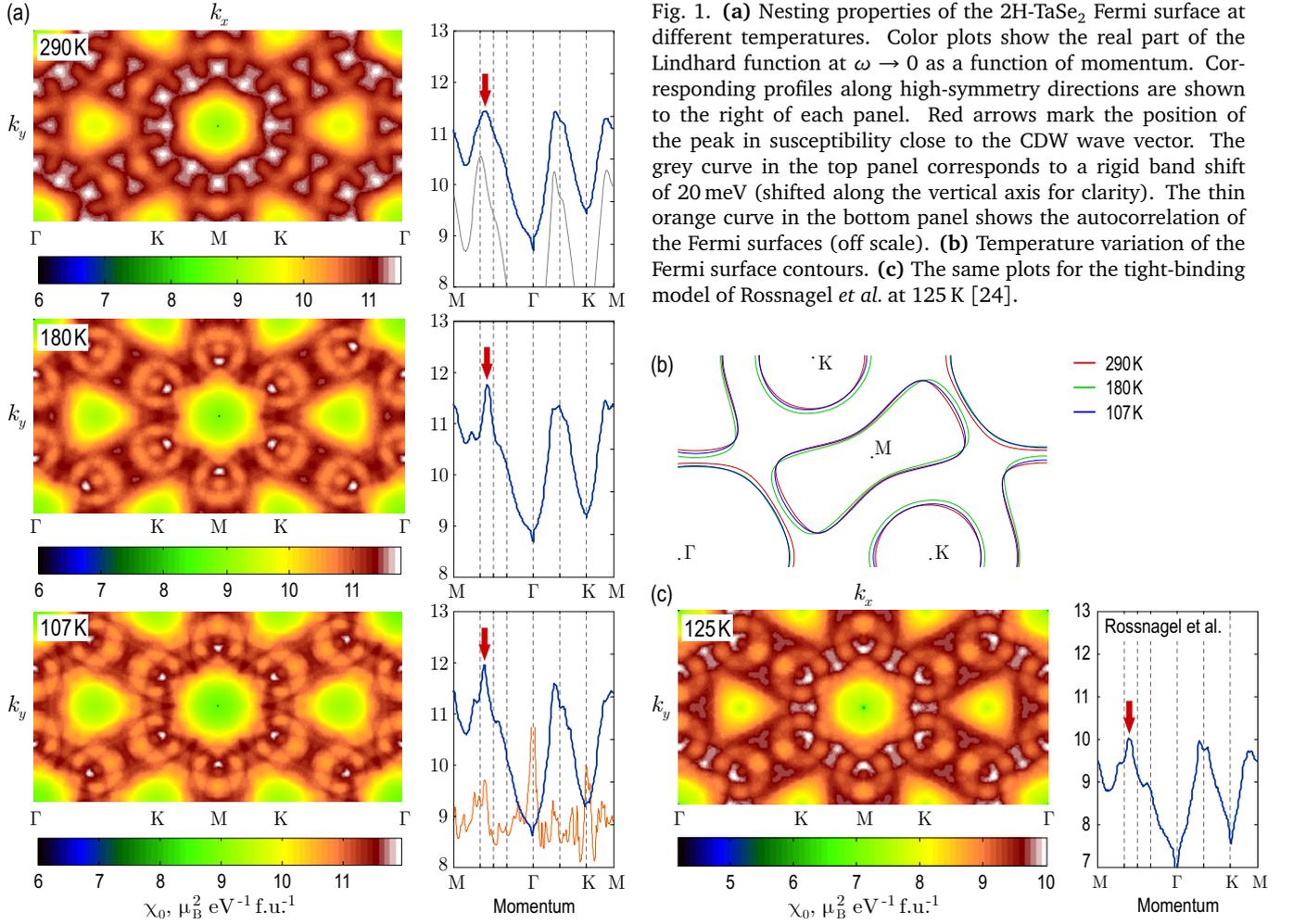

Fig. 1. **(a)** Nesting properties of the 2H-TaSe$_2$ Fermi surface at different temperatures. Color plots show the real part of the Lindhard function at $\omega \to 0$ as a function of momentum. Corresponding profiles along high-symmetry directions are shown to the right of each panel. Red arrows mark the position of the peak in susceptibility close to the CDW wave vector. The grey curve in the top panel corresponds to a rigid band shift of 20 meV (shifted along the vertical axis for clarity). The thin orange curve in the bottom panel shows the autocorrelation of the Fermi surfaces (off scale). **(b)** Temperature variation of the Fermi surface contours. **(c)** The same plots for the tight-binding model of Rossnagel *et al.* at 125 K [24].

### Nesting properties of 2H-TaSe$_2$ as a function of temperature

2H-TaSe$_2$ (trigonal prismatic tantalum diselenide) is a quasi-two-dimensional CDW-bearing material with two phase transitions at accessible temperatures: a second-order incommensurate CDW transition at 122 K and a first-order commensurate 3×3 CDW lock-in transition at 90 K [3]. The temperature evolution of its Fermi surface can be found for instance in Ref. 19. The Fermi surface sheets originate from two bands: one is responsible for the Γ and K barrels with a saddle point in between, the other one supports the "dogbone" with another saddle point at M. The dispersions in the normal and incommensurate CDW states are qualitatively similar [19, 24, 32, 33]. In contrast, the lock-in transition to the commensurate CDW state at 90 K is clearly pronounced, resulting in a new folded Fermi surface consisting of a set of nearly circles around new Γ points and rounded triangles around new K points [19]. The commensurate CDW vectors $q_n = \frac{2}{3}\Gamma M$ are well known from experiments [2, 3].

The question that we want to address here is whether the Fermi surface geometry observed in the normal state by ARPES possesses the nesting properties that could explain the transition to the CDW state upon cooling. It is also interesting to which extent the Fermi surface varies with temperature in the neighborhood of the incommensurate CDW transition. To our knowledge, in earlier studies such minor variations of the Fermi surface could not be detected, and the dispersion was considered unchanged down to 90 K [24, 32, 33], until the Fermi surface reconstruction due to the commensurate CDW lock-in transition finally occurred. On the other hand, we have recently reported a noticeable variation of the distance between the M- and K-barrels (see Fig. 4a in Ref. 19), which gave us the motivation to study temperature variations of the Fermi surface all over the momentum space in order to estimate their effect on the nesting properties.

| | Γ- and K-barrels | | | | | M-barrels ("dogbones") | | | | |
|---|---|---|---|---|---|---|---|---|---|---|
| $T$ (K) | $t_0^a$ | $t_1^a$ | $t_2^a$ | $t_3^a$ | $t_4^a$ | $t_0^b$ | $t_1^b$ | $t_2^b$ | $t_3^b$ | $t_4^b$ |
| 290 | −0.027 | 0.199 | 0.221 | 0.028 | 0.013 | 0.407 | 0.114 | 0.444 | −0.033 | 0.011 |
| 180 | −0.051 | 0.172 | 0.248 | 0.005 | 0.011 | 0.355 | −0.015 | 0.406 | −0.069 | 0.013 |
| 107 | −0.064 | 0.167 | 0.211 | 0.005 | 0.003 | 0.369 | 0.074 | 0.425 | −0.049 | 0.018 |

Table I. Experimental tight-binding parameters of 2H-TaSe$_2$ independently determined for three different temperatures. All values are given in eV.

We have fitted the experimental dispersion of 2H-TaSe$_2$ measured with high resolution at three different temperatures using the following tight-binding expansion (momenta $k_x$ and $k_y$ enter the formula in dimensionless units):

$$\epsilon_\mathbf{k} = t_0 + t_1 \left[ 2\cos\frac{k_x}{2}\cos\frac{\sqrt{3}\,k_y}{2} + \cos k_x \right] + t_2 \left[ 2\cos\frac{3k_x}{2}\cos\frac{\sqrt{3}\,k_y}{2} + \cos\sqrt{3}\,k_y \right] \\ + t_3 \left[ 2\cos k_x \cos\sqrt{3}\,k_y + \cos 2k_x \right] + t_4 \left[ 2\cos 3k_x \cos\sqrt{3}\,k_y + \cos 2\sqrt{3}\,k_y \right]. \quad (2)$$

The tight-binding parameters were independent for the two bands, which resulted in the total of 10 fitting parameters. The fitting was done by precisely measuring the relative Fermi momenta (distances between Fermi surface contours) and Fermi velocities along several high-symmetry directions that were exactly determined from Fermi surface maps at different temperatures. The tight-binding parameters were then found by solving an overdetermined system of 15 equations relating the Fermi momenta and velocities of the model to the experimentally measured ones, so that the resulting tight-binding model best reproduces both the Fermi surface contours and the experimental dispersion in the vicinity of the Fermi level. Such fitting procedure has been applied to the Fermi surface maps of 2H-TaSe$_2$ independently at three temperatures: 107 K (in the incommensurate CDW state), 180 K, and 290 K (both in the normal state). The corresponding tight-binding parameters are given in Table I, and the Fermi surface contours are shown in Fig. 1 (b) for comparison.

We then calculated the Lindhard functions at $\omega \to 0$ as

$$\chi_\mathbf{q} = \sum_\mathbf{k} \frac{n_\text{F}(\epsilon^\text{a}_\mathbf{k}) - n_\text{F}(\epsilon^\text{a}_\mathbf{k+q})}{\epsilon^\text{a}_\mathbf{k} - \epsilon^\text{a}_\mathbf{k+q}} + \sum_\mathbf{k} \frac{n_\text{F}(\epsilon^\text{a}_\mathbf{k}) - n_\text{F}(\epsilon^\text{b}_\mathbf{k+q})}{\epsilon^\text{a}_\mathbf{k} - \epsilon^\text{b}_\mathbf{k+q}} + \sum_\mathbf{k} \frac{n_\text{F}(\epsilon^\text{b}_\mathbf{k}) - n_\text{F}(\epsilon^\text{a}_\mathbf{k+q})}{\epsilon^\text{b}_\mathbf{k} - \epsilon^\text{a}_\mathbf{k+q}} + \sum_\mathbf{k} \frac{n_\text{F}(\epsilon^\text{b}_\mathbf{k}) - n_\text{F}(\epsilon^\text{b}_\mathbf{k+q})}{\epsilon^\text{b}_\mathbf{k} - \epsilon^\text{b}_\mathbf{k+q}}, \quad (3)$$

where indices a and b indicate the two bands forming the Γ- and K-centered hole barrels and M-centered electron "dog-bones" respectively. The results of the calculation are shown in Fig. 1 (a). The sharp peak seen near the $\frac{2}{3}$ΓM wave vector (red arrows) is a clear evidence of nesting. Surprisingly, it does not exactly coincide with the CDW vector, but appears at ∼ 0.58–0.60 ΓM. The same calculation performed for the tight-binding model of Rossnagel *et al.* [24], as shown in Fig. 1 (c), yields the same pattern of somewhat weaker peaks at remarkably similar positions. As will be shown later, similar incommensurate nesting peak appears to be universal between different transition metal dichalcogenides.

The temperature behavior of the nesting vector observed in TaSe$_2$ agrees with our previous observations [19]. Upon lowering the temperature towards the incommensurate CDW transition, the nesting vector moves away from the commensurate position, which means that the system feels the instability and starts to avoid it already above the transition. In the incommensurate state, the nesting peak seems to be slightly driven in the opposite direction upon cooling, which finally drives the commensurate transition at 90 K.

As also seen from Fig. 1, the absolute intensity of the dominant nesting peak slightly decreases with temperature due to natural temperature broadening, which finally leads to the phase transition as soon as the instability criterion (1) is satisfied. This natural scenario is confirmed by the observation of a Kohn-like anomaly in the $\Sigma_1$ phonon branch already at 300 K, which softens even more as the transition is approached [9]. Such a mutual response can signify a strong electron-phonon interaction in 2H-TaSe$_2$. It is interesting that at first the system does not develop a static commensurate CDW order. Instead, it opens up a pseudogap [19] and falls into an incommensurate CDW state, which shifts the nesting vector closer to the commensurate position [see the 107 K curve in Fig. 1 (a)] preserving its strength. This new nesting peak in the incommensurate state may finally drive the commensurate CDW transition at lower temperatures.

We note here that the effect of temperature on the absolute value of the susceptibility may be even higher due to the renormalization effects, as the self-energy is usually temperature-dependent. Therefore many-body effects, which we neglect in our calculations, may lead to additional temperature broadening of the spectral function and consequently of the susceptibility.

Our results are at variance with the previous observations [18, 24, 26], which have found susceptibility for 2H polytypes to take a broadly humped form without strong signatures of nesting, and with the earlier band structure calculations [25], which fail to reproduce the Fermi surface topology and therefore its nesting properties.

It is worth mentioning that in our previous work [19] the calculation of the Lindhard function as described above was replaced by the autocorrelation of the Fermi surface maps, which is easier to calculate. This procedure is partially justified, as the peaks that are present in the Lindhard function are also to be found in the autocorrelation. One has to be careful, however, as the latter may additionally include many "false" peaks that do not represent relevant nesting vectors [compare two curves in the bottom panel of Fig. 1 (a)]. The rigorous calculation of the bare susceptibility should be therefore preferred whenever allowed by the computational capability.

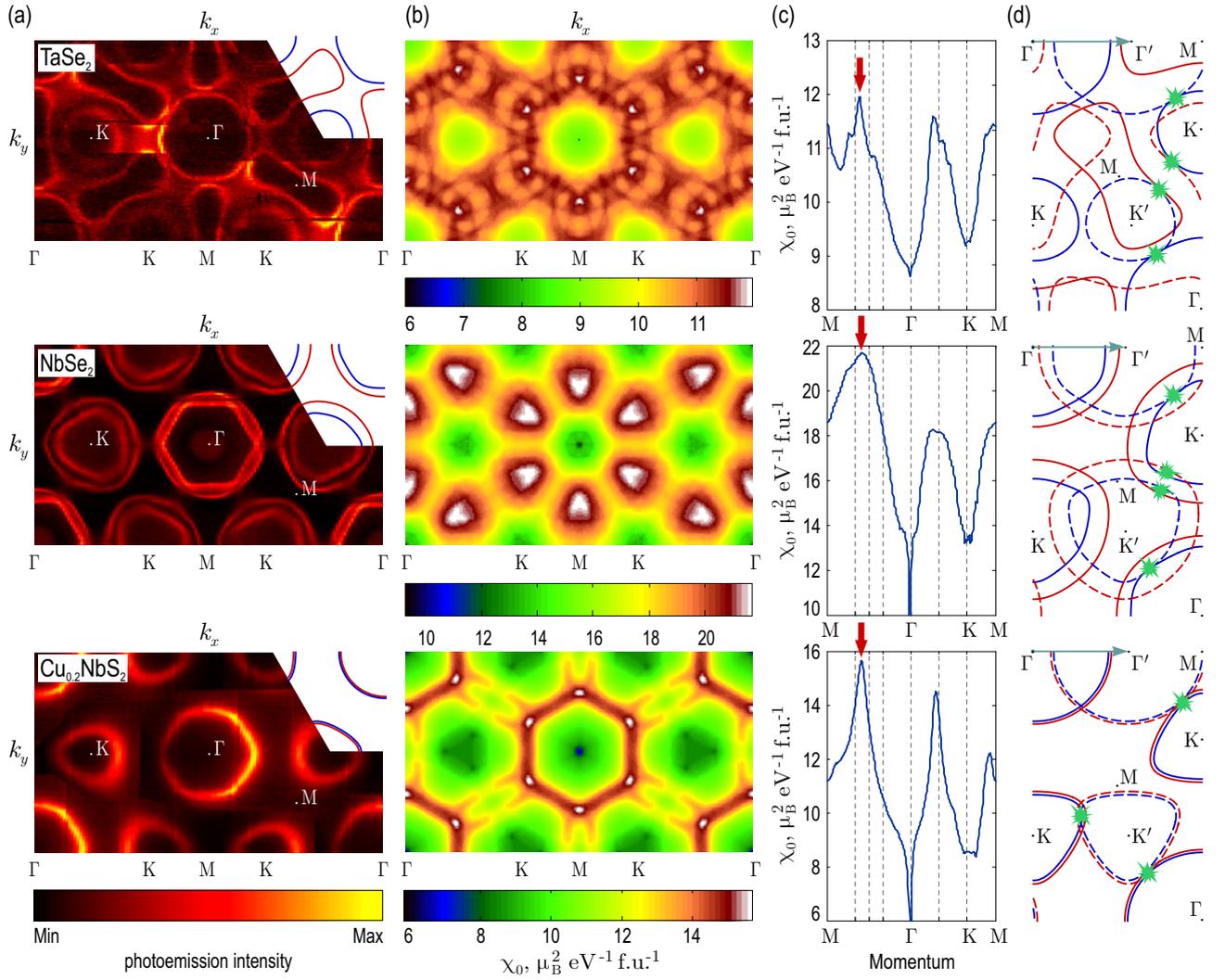

Fig. 2. Nesting properties of three transition metal dichalcogenides: 2H-TaSe$_2$ (top row), 2H-NbSe$_2$ (middle row), and 2H-Cu$_{0.2}$NbS$_2$ (bottom row). **(a)** Experimental normal-state Fermi surfaces as seen by ARPES. The maps are symmetrized so that they fully cover the rectangular unitary cell in the momentum space. **(b)** Real part of the Lindhard function at $\omega \to 0$ as a function of momentum. **(c)** Corresponding profiles along high-symmetry directions, with the dominant nesting vector marked by the red arrow. The same vectors can be seen in panel (b) as white spots. **(d)** Fermi surface contours (solid lines) are shown together with their replicas shifted by the nesting vector (dashed lines) to demonstrate the nesting geometry. The nesting vector is shown on top of each image. The parts of the Fermi surface that give most contribution to the susceptibility are marked by green "sparks".

**Universality of the nesting vector**

For a comparison with TaSe$_2$, we have chosen two other transition metal dichalcogenides of the same 2H polytype, namely NbSe$_2$ (niobium diselenide) and Cu$_{0.2}$NbS$_2$ (copper intercalated niobium disulphide). Both compounds possess different Fermi surface topology, as in contrast to 2H-TaSe$_2$, both saddle points along the $\Gamma$K line are located below the Fermi level, therefore the Fermi surface consists of double hole barrels around the $\Gamma$ and K points, rather than single hole barrels and single electron "dogbones" as in case of TaSe$_2$. The experimental Fermi surfaces of all three compounds are presented in Fig. 2 (a), measured at 180 K, 20 K, and 30 K respectively. Both TaSe$_2$ and NbSe$_2$ are half-filled band metals, while copper intercalation in Cu$_{0.2}$NbS$_2$ is responsible for the 20% electron doping. These doping levels are in good agreement with Fermi surface areas determined from ARPES datasets.

The ordering temperatures of the three materials are also different. 2H-TaSe$_2$ experiences an incommensurate CDW phase transition at 122 K, commensurate CDW phase transition at 90 K, and a superconducting transition at 0.2 K [34]. In 2H-NbSe$_2$ incommensurate CDW transition happens at much lower temperature of 33.5 K [2, 3, 35], the commensurate CDW transition does not occur, whereas the superconducting critical temperature is enhanced to 7.2 K [5]. 2H-NbS$_2$ shows no CDW transitions (neither incommensurate nor commensurate) [36], but to our knowledge it has not been studied whether CDW ordering exists in Cu-intercalated samples around 20% doping. Superconductivity is still present below 6.3 K in pure NbS$_2$ [34], but is suppressed to 2.5 K in Cu$_{0.2}$NbS$_2$.

One would expect the nesting properties in these three materials to be different, the absolute value of the susceptibility at the dominant nesting vector being in direct correspondence to the CDW transition temperature. Surprisingly, the results of our calculations shown in Fig. 2 (b) reveal a totally different picture. First, the maximal value of the susceptibility is not correlated with the CDW transition temperature. It is the lowest among the three compounds in TaSe$_2$, where $T_{\text{CDW}}$ is the highest, is higher in Cu$_{0.2}$NbS$_2$, where no CDW order has been observed, and is maximal in NbSe$_2$, which has an intermediate transition temperature. This probably could be explained by the difference in the phonon spectra and electron-phonon interaction in these materials or by the differences in $k_z$-dispersion, which we neglect in our calculations. More surprising is that the dominant nesting vector in both compounds coincides with that of TaSe$_2$ and is located at 0.60±0.05 ΓM [see panel (c) of the same figure]. Such coincidence can hardly be accidental, as we know that the nesting properties are extremely sensitive to the Fermi surface geometry. In fact, a rigid band shift of the TaSe$_2$ band structure by 20 meV is enough to displace the nesting peak in the room temperature susceptibility from 0.6 ΓM to the commensurate position at $\frac{2}{3}$ΓM. Such shift corresponds to the electron doping of about 4%, which is much smaller than the 20% doping of Cu$_{0.2}$NbS$_2$. Similarly, an arbitrary distortion of the Fermi surface in any of these compounds was shown to destroy the universal nesting vector or shift it to a different position.

|  | inner barrels | | | | | outer barrels | | | | |
|---|---|---|---|---|---|---|---|---|---|---|
|  | $t_0^a$ | $t_1^a$ | $t_2^a$ | $t_3^a$ | $t_4^a$ | $t_0^b$ | $t_1^b$ | $t_2^b$ | $t_3^b$ | $t_4^b$ |
| 2H-NbSe$_2$ | 0.0003 | 0.0824 | 0.1667 | 0.0438 | 0.0158 | 0.1731 | 0.1014 | 0.2268 | 0.0370 | –0.0048 |
| 2H-Cu$_{0.2}$NbS$_2$ | –0.029 | 0.191 | 0.235 | 0.108 | 0.000 | 0.011 | 0.196 | 0.230 | 0.098 | 0.000 |

Table II. Experimental tight-binding parameters of 2H-NbSe$_2$ and 2H-Cu$_{0.2}$NbS$_2$. All values are given in eV.

The tight binding parameters that were fitted to experimental ARPES datasets and used for susceptibility calculations are given in Table II. The corresponding Fermi surfaces for all three materials are shown in Fig. 2 (d) together with their replicas shifted by the nesting vector (dashed lines) to show the different nesting geometry. In TaSe$_2$ the dominant contribution to the nesting peak comes from simultaneous tangency of the M and K barrels, K and Γ barrels, and the K barrel with itself. In NbSe$_2$ the broad nesting peak is a sum of several peaks that originate from the pairwise tangency of the two K-centered barrels with the Γ barrels and with themselves, while in Cu$_{0.2}$NbS$_2$ the splitting between the two bands is too small to be resolved, so they can be thought of as a single degenerate band that produces the sharp nesting peak by the external contact of the K-centered barrel with the Γ barrel and with itself, as shown in the figure. Such different nesting geometries make the coincidence of their nesting vectors even more puzzling.

**Conclusions and discussion**

To conclude, we have found that the Fermi surfaces of TaSe$_2$, NbSe$_2$, and Cu$_{0.2}$NbS$_2$ possess a strong nesting vector in the vicinity of the CDW wave vector, which supports the Peierls instability scenario for the formation of CDW/PLD in these materials. We observe however an offset of the nesting vector from the commensurate position that is persistent over Fermi surfaces of all three materials and over different temperatures in TaSe$_2$, and note that the absolute maximal value of the susceptibility does not correlate with the ordering temperature.

Here several questions can be posed. First, how to explain the inconsistency of the maximal susceptibility values with the transition temperature. As one would suspect that the differences in the electron-phonon coupling might be a possible answer, a comparative study of the phonon spectrum in these materials seems to be necessary. Second, why is the nesting vector shifted from the commensurate position, if we know that even in the incommensurate CDW state the incommensurability of the CDW wave vector does not exceed 2% [3]? A small shift of the CDW wave vector relative to the nesting vector is natural, if one recalls that in the parent compound the frequency of the phonon mode is nonzero, which means that it will couple to the susceptibility at a finite energy of the phonon mode (which is of the order of 10 meV), rather than at the Fermi level. Evidently, the CDW does not necessarily form exactly at the same wave vector at which the instability criterion (1) is first satisfied (a similar discrepancy in pure Cr and its possible mechanisms are discussed for example in Ref. 37). In fact, Eq. (1) provides the conditions at which the CDW transition becomes energetically favorable, but does not specify the exact lattice configuration, at which the new energy minimum is reached. As the phonon frequency softens towards zero, it will couple to the electronic susceptibility at different energies, which correspond to slightly different positions of the nesting peak in momentum space. Moreover, the electronic system itself may react to the ongoing transition, changing its nesting vector. Slight deviations of the CDW wave vector from the peak in susceptibility would also be possible if the electron-phonon interaction is strongly momentum-dependent, which is however an unlikely explanation, as the nesting peaks, at least in TaSe$_2$ and Cu$_{0.2}$NbS$_2$, are very sharp.

Finally, we note that the universality of the nesting vector among different compounds is in line with the neutron scattering measurements performed on the same materials [3], which show surprisingly identical incommensurate wave vectors in 2H-TaSe$_2$ and 2H-NbSe$_2$. The fact that the nesting vector in Cu$_{0.2}$NbS$_2$ is the same possibly means that Cu-intercalation enhances the CDW instability is this material, even though no CDW order is observed in the pure compound. The presence of CDW order in Cu$_{0.2}$NbS$_2$ would not be surprising, being analogous to the strong changes in the magnetic transition temperatures observed in NbS$_2$ upon intercalation by the first row transition metals (Mn, Fe, Co, Ni) [38, 39] to comparable doping levels. Alternatively, our findings might suggest that chemical intercalation does not result in a simple rigid band shift of the bands, which would immediately destroy the nesting vector, but leads to more complex changes in the dispersion that pin some parts of the Fermi surface relevant for the nesting to their original position. A similar effect has been observed by C. Battaglia *et al.* [40], who have shown that the Γ-centered barrel in Ni- and Mn-intercalated NbS$_2$ remains practically unaffected by the presence of intercalant species in violation of the rigid band approximation. To distinguish between these two possibilities, more systematic studies of the nesting properties as a function of doping and in other transition metal dichalcogenides might be helpful.

## Acknowledgements

This project is part of the Forschergruppe FOR538 and is supported by the DFG under Grants No. KN393/4 and BO1912/2-1. The work in Lausanne was supported by the Swiss National Science Foundation and by the MaNEP. ARPES experiments were performed using the 1$^3$ ARPES end station at the UE112-lowE PGMa beamline of the Berliner Elektronenspeicherring-Gesellschaft für Synchrotron Strahlung m.b.H. (BESSY). We want to thank Dr. Bussy for carrying out the chemical analysis of the Cu$_{0.2}$NbS$_2$ single crystals. This work has been supported by the Swiss National Foundation for the Scientific Research within the NCCR MaNEP pool. We also thank Ch. Heß and A. Kondrat for resistivity measurements and R. Hübel for technical support.